\documentclass[11pt]{article}
\usepackage[utf8]{inputenc}
\usepackage[T1]{fontenc}
\usepackage{lmodern}
\usepackage{textcomp}
\DeclareUnicodeCharacter{2212}{\ensuremath{-}}
\DeclareUnicodeCharacter{2264}{\ensuremath{\le}}
\DeclareUnicodeCharacter{2265}{\ensuremath{\ge}}
\DeclareUnicodeCharacter{03B1}{\ensuremath{\alpha}}
\DeclareUnicodeCharacter{03C3}{\ensuremath{\sigma}}
\DeclareUnicodeCharacter{03A3}{\ensuremath{\Sigma}}
\DeclareUnicodeCharacter{03C4}{\ensuremath{\tau}}
\DeclareUnicodeCharacter{221A}{\ensuremath{\surd}}
\DeclareUnicodeCharacter{2248}{\ensuremath{\approx}}
\DeclareUnicodeCharacter{00D7}{\ensuremath{\times}}
\DeclareUnicodeCharacter{00B7}{\ensuremath{\cdot}}
\DeclareUnicodeCharacter{00B2}{\ensuremath{^2}}
\usepackage[margin=2.5cm]{geometry}
\usepackage{booktabs}
\usepackage{amsmath}
\usepackage{graphicx}
\usepackage[hidelinks]{hyperref}
\usepackage{longtable}
\usepackage{array}
\usepackage{calc}

\hypersetup{pdftitle={Zero-shot time-series foundation models against the Khepri v1.5 baselines for 96-hour carbon-intensity forecasts in nine Nordic bidding zones: a preregistered comparison}, pdfauthor={Eirik Botten Nicolaysen}}

\title{Zero-shot time-series foundation models against the Khepri v1.5 baselines for 96-hour
carbon-intensity forecasts in nine Nordic bidding zones: a preregistered comparison}
\author{Eirik Botten Nicolaysen\\EcoDeco AS, Norway\\\texttt{eirik@ecodeco.no}}
\date{2026-11-02}

\begin{document}
\sloppy\setlength{\emergencystretch}{3em}
\maketitle

\begin{abstract}
We compare three zero-shot time-series foundation models (Timestral-5M, Timestral-209M and
Chronos-2, 8192-hour context, no covariates) with the Khepri v1.5 SARIMA and GBM baselines,
with their code and hyperparameters unchanged, for 96-hour forecasts of hourly carbon intensity in the Nordic bidding zones NO1--NO5 and SE1--SE4. The
design, the analysis code and a minimum detectable effect (MDE) were timestamped before any
foundation-model forecast by an OpenTimestamps proof (Bitcoin-confirmed) and the GPG-signed tag
prereg-v1; the proofs are archived in Zenodo (DOI 10.5281/zenodo.23881315). The preregistration was also
submitted to OSF Registries 2026-11-02 07:49 UTC (osf.io/hncz5; embargo requested, pending OSF approval). On the 2025 test (361 daily origins,
contamination not ruled out) all three models have a lower zone-averaged MAPE than both baselines,
by 0.952 to 1.347 percentage points (Diebold--Mariano with the Harvey--Leybourne--Newbold
correction, all $p<0.0001$), and each beats SARIMA in 9 of 9 zones. On a post-release holdout of
14 origins from 2026-10-16 the point estimates have the same sign (0.729 to 1.326 percentage
points lower), but all lie within the preregistered minimum detectable effect of 2.989 points and
are reported as not determined. For calibration (RQ4) the preregistration locked a Diebold--Mariano--HLN test on CRPS;
the MDE applies to MAPE only. Holdout CRPS point estimates (n = 14 origins) are
1.1114 (Timestral-5M), 1.0733 (Timestral-209M) and 1.1198 (Chronos-2) against 1.4165 gCO$_2$-eq/kWh
for SARIMA ($p<0.0001$ for each comparison). The SARIMA quantiles are constructed from the fitted
model's Gaussian predictive distribution (deviation D-03), not taken from a native v1.5 forecast, so the
difference may partly reflect the SARIMA intervals. The contamination indicator does not exceed its
preregistered threshold for any comparison; with 14 holdout origins it has low power (HAC
standard errors 0.124 to 0.360 pp). Forecasts, the CI series,
the locked analysis and all deviations are released with the code.
\end{abstract}

\section{Introduction}\label{introduction}

Forecasts of the carbon intensity (CI) of electricity, in grams of CO2-equivalent per kWh, have been
proposed as a building block for shifting flexible electricity use to hours with lower
emissions (Maji et al.~2022a).

For the nine Nordic bidding zones NO1--NO5 and SE1--SE4, Khepri v1.5 computes an hourly CI series
from ENTSO-E generation per production type and evaluates SARIMA and gradient-boosting (GBM)
baselines over a 96-hour horizon from daily origins (Khepri v1.5, DOI 10.5281/zenodo.23101127).

On the 2025 test, GBM beat SARIMA in 7 of the 9 zones, by 0.02 to 2.64 percentage points (pp) of
mean absolute percentage error (MAPE), and lost in NO4 (+2.92 pp) and NO5 (+0.53 pp).

Time-series foundation models (TSFMs) are pretrained forecasters that can be applied to a new
series without training on it (``zero-shot'').
Chronos-2 is described by its developers as
offering zero-shot support for univariate, multivariate and covariate-informed forecasting
(Ansari et al.~2025). Timestral, released by Mistral AI on 2026-10-15 (Mistral AI 2026), comes in four sizes
from 5M to 209M parameters with 21 predictive quantiles per time step.

For a CI series, a zero-shot TSFM needs only the history of the series itself. This makes it
directly comparable with the Khepri baselines, which are also univariate and fitted per zone.

TSFMs have already been applied to CI: the public CarbonX repository lists zero-shot forecasts
with five TSFMs for 214 grids, among them Norway and Sweden as whole countries (CarbonX).

At bidding-zone level the baselines differ between zones: in v1.5, NO5 had by
far the lowest MAPE level (SARIMA 0.72 \%), and NO4 and NO5 were the two zones where GBM lost to
SARIMA.

Evaluating a pretrained model on historical data raises a contamination problem. If the test
period, or the series itself, was part of the pretraining corpus, a retrospective test can
overstate the accuracy the model would have on data it has not seen. The repository of
the live benchmark Impermanent states the same concern for static benchmarks: ``{[}t{]}raining data may
overlap with evaluation data'' (Garza et al.~2026).

For the models used here the corpus cannot be checked directly. The Timestral repository
discloses neither training data nor a training cut-off. For Chronos-2 the pretraining datasets
are named on its model card; in our reading of the card no dataset name refers to ENTSO-E, Norway
or Sweden, but a name match cannot rule out that Nordic data entered through aggregated datasets.

We therefore pair two tests. A retrospective test on the Khepri v1.5 2025 test set (361 daily
origins per zone) has more statistical power but is labelled ``contamination not ruled out''. A
post-release holdout uses only origins from 2026-10-16, the day after the Timestral release, so
that every target hour lies after the release of all three TSFMs; it has 14 origins per zone.

Fourteen origins give little power. The minimum detectable effect (MDE) for the zone-averaged MAPE
difference, estimated from the SARIMA − GBM loss difference on the 2025 test, is 2.989 pp on the holdout, against 0.413 pp on the retro test. A holdout difference smaller than
this is reported as ``not determined'', not as equivalence.

The holdout is therefore a contamination control, and the difference between the TSFM advantage
on the two tests is reported as a descriptive contamination indicator, not as a test.

The contributions are as follows. We make no claim of priority.

\begin{itemize}
\item
  A preregistered comparison of three zero-shot TSFMs (Timestral-5M, Timestral-209M,
  Chronos-2) with the Khepri v1.5 SARIMA and GBM baselines and a seasonal-naive reference, for
  96-hour CI forecasts in nine Nordic bidding zones. Design, analysis code and MDE were
  timestamped before any foundation-model forecast by an OpenTimestamps proof (Bitcoin-confirmed)
  and the GPG-signed tag \texttt{prereg-v1}; the proofs are archived in Zenodo (10.5281/zenodo.23881315). The
  preregistration was also submitted to OSF Registries 2026-11-02 07:49 UTC (osf.io/hncz5; embargo requested, pending OSF approval).
\item
  A design that pairs a retrospective test with a small minimum detectable effect and a
  post-release holdout, with the minimum detectable effect and the decision rule fixed before
  any TSFM forecast was made.
\item
  Hourly predictions of every arm, the derived CI series and the code, stored with sha256 sums
  and OpenTimestamps stamps, and a deviation log in which every deviation is dated.
\item
  A prospective log that issues all arms daily from the registration onward, to be analysed
  later with the same locked analysis code.
\end{itemize}

The rest of the paper is organised as follows. Section 2 reviews related work, Section 3
describes the data, arms, tests and deviations, Section 4 reports the results and Section 5
describes the software and its availability.

\section{Related work}\label{related-work}

\subsection{Carbon-intensity forecasting}\label{carbon-intensity-forecasting}

DACF (Maji et al.~2022a) forecasts the hourly average CI of a grid for the next 24 hours. It
predicts the production of each generation source with a separate neural network that uses
historical production, calendar features and day-ahead weather forecasts, and combines the
source forecasts with direct (operational) emission rates.

It is evaluated on six regions, four in the US and two in Europe, Sweden (SE) and Germany (DE),
with ENTSO-E data for the European regions. It uses data from 2019--2021, retrains its models every
six months and reports predictions for the last six months of 2021 as the test, with MAPE as the
metric.

DACF is compared with the authors' own reimplementation of the method of Leerbeck et al.~(2020),
since that code was not public.

According to its public repository, CarbonCast (Maji et al.~2022b) extends DACF to
forecasts of up to 96 hours. The repository covers 13 regions, among them Sweden (SE) as the only
Nordic region, and its saved models are trained on 2020 and validated on January--June 2021; the
repository states that results similar to the paper are obtained when testing on July--December
2021. It also points to a real-time v3.0 branch in testing, which we did
not read.

The repository of EnsembleCI (Yan et al.~2025) implements CI forecasting with
gradient-boosting, neural and ensemble learners (LightGBM, a FastAI network, CatBoost, XGBoost and
an AutoGluon weighted ensemble) and lists 11 regions, with Sweden (SE) as the only Nordic one.

In the region lists and released data of these three repositories, Sweden appears only at
country level and Norway does not appear; no series for the bidding zones NO1--NO5 or SE1--SE4 is
included.

Leerbeck et al.~(2020) forecast average and marginal CO2 emission intensity for the Danish bidding
zone DK2 up to 24 hours ahead with 95 \% prediction intervals, combining linear models with
exogenous inputs (ARIMAX) fitted to data supplied by Tmrow.

The EnsembleCI paper (Yan et al.~2025) evaluates a stacked ensemble of gradient-boosting and neural
learners on 11 grids, six in the US and five in the EU with Sweden as a country, up to four days
ahead with MAPE, testing the EU grids on the second half of 2021, and reports lower MAPE than
CarbonCast on all four days.

Clauß et al.~(2019) compute hourly average CO2-equivalent intensities for Scandinavian bidding
zones with an input--output method that accounts for trade between zones, from ENTSO-E generation
data (Swedish zones from the Swedish TSO), and use them as a control signal for the demand response
of residential heating; the intensities are computed for 2015 and are not forecast.

GEFCom2014 had four probabilistic forecasting tracks, on load, price, wind and solar power (Hong
et al.~2016).

\subsection{Foundation models for carbon intensity}\label{foundation-models-for-carbon-intensity}

The CarbonX paper (Maji et al.~2025) evaluates five TSFMs (MOMENT, Chronos, TimesFM, Sundial and
Time-MoE) zero-shot for CI forecasts up to 96 hours, extended to 21 days. Its benchmark has 13 grids;
the European ones are Sweden (SE), Germany, Poland, Spain and the Netherlands, evaluated on the last
six months of 2021, and a global evaluation uses Electricity Maps data for 214 grids (2021--2024,
70/30 split) whose names are not given in the paper. The evaluations are on historical data; we found
no discussion in the paper of overlap between the models' pretraining data and the evaluation data,
and no Norwegian grid or bidding zone is named.

The CarbonX repository provides zero-shot CI forecasts with five TSFMs (Chronos, MOMENT, Sundial,
Time-MoE and TimesFM) for 214 grids at a horizon of up to 96 hours, with an extended mode of up to
21 days.

Norway and Sweden are included as whole countries; the released data for both are hourly series
from 2021-01-01 to 2024-12-31, and no bidding-zone series is released.

We found no statement in the repository about prospective evaluation, evaluation on data after
the models' release, or overlap between the models' pretraining data and the evaluation data.

Whether the repository corresponds exactly to the evaluation reported in the CarbonX paper could
not be verified.

The three TSFMs evaluated here come from two families: Timestral (Mistral AI) and Chronos-2
(Ansari et al.~2025). The Chronos-2 technical report describes a pretrained model for univariate,
multivariate and covariate-informed forecasting in a zero-shot manner; its context length is
extended to 8192 in a second training stage, multivariate and covariate-informed training relies
entirely on synthetic data, and none of the real univariate pretraining datasets it lists (its
Table 6) is an ENTSO-E or CI series; the energy-domain datasets listed are Electricity, Solar, Wind
Farms, London Smart Meters and Buildings 900K. The report gives no training cut-off or release date.

The Timestral code is released under Apache-2.0 and the weights separately under a
``Modified MIT License''; its repository discloses neither training data nor a training cut-off.
Chronos-2 became public in October 2025: its technical report appeared on 2025-10-17, its
repository announces the release on 2025-10-20, and the Hugging Face model repository
amazon/chronos-2 dates from 2025-10-30 (D-12). The quickstart notebook in its repository gives a
maximum context of 8192 steps.

\subsection{Contamination and live evaluation}\label{contamination-and-live-evaluation}

According to its repository, Impermanent (Garza et al.~2026) is a live benchmark for
temporal generalisation in which models must forecast before the future is observed and are
scored as data arrive. The repository motivates this with the risk that, for foundation models,
``{[}t{]}raining data may overlap with evaluation data'' in static benchmarks. The signals it lists are
activity series derived from GitHub Archive (stars, pushes, pull requests, issues).

Our post-release holdout is not a live evaluation: its forecasts were produced after the truth
had been observed. It guarantees only that the target hours postdate the public release of
Timestral (2026-10-15, the date of its GitHub release commit) and of Chronos-2 (October 2025). In
the prospective log (Section 3.10) a scheduled job issues the forecast for each 00:00 UTC origin
a few hours later, while most of the 96 target hours are still unobserved; the context ends at the
origin.

In the Impermanent paper, every model must produce point and probabilistic forecasts for the next horizon at
each cutoff before any ground truth is available, and forecasts are scored after the outcomes
arrive. The paper evaluates four TSFMs, among them Chronos-2, on GitHub activity series from GH Archive and
contains no energy or CI series.

\subsection{Forecast comparison and evaluation}\label{forecast-comparison-and-evaluation}

Real-time probabilistic forecasting exercises outside energy evaluate forecasts submitted as
quantiles with the weighted interval score and the empirical coverage of prediction intervals, for
example the US COVID-19 Forecast Hub, whose forecasts were submitted as 23 quantiles from 0.01 to
0.99 (Cramer et al.~2022).

We compare forecast accuracy with the Diebold--Mariano test (Diebold \& Mariano 1995) with the
small-sample correction of Harvey, Leybourne \& Newbold (1997), and evaluate predictive
quantiles with the CRPS (Gneiting \& Raftery 2007); details are in Section 3.

\section{Methods}\label{methods}

\subsection{Study design and preregistration}\label{study-design-and-preregistration}

The design, the analysis code and deviations D-01 to D-08 were fixed in the tagged
preregistration before any time-series foundation model (TSFM) forecast was produced on real
carbon-intensity (CI) data; D-09 to D-12 came later (Section 3.9).

The design, the analysis code and the MDE were timestamped before any foundation-model forecast
by an OpenTimestamps proof (Bitcoin-confirmed) of the tarball of the repository at the GPG-signed
git tag \texttt{prereg-v1} (commit 74539f2af7bb6d203d8c8452e01d96a222156078) and by the signed tag itself;
the proofs are archived in Zenodo (10.5281/zenodo.23881315).

The preregistration was also submitted to OSF Registries 2026-11-02 07:49 UTC (osf.io/hncz5; embargo requested, pending OSF approval).

The locked analysis is \texttt{analysis.py} with \texttt{khepri\_tsfm/stats.py} at that tag.

Deviations are listed in Section 3.9 and in full in the repository
(\texttt{truth/DEVIATIONS\_KHEPRI-TSFM\_v1\_2026-11-02.md}).

The study has two test windows with different roles. A \emph{retrospective} test (retro) on the
Khepri v1.5 2025 test set carries the statistical power but is labelled ``contamination not
ruled out'': the training cut-off of Timestral is not disclosed in its repository, and most of the 2025 test period
precedes the public release of Chronos-2 in October 2025 (D-12). A \emph{post-release holdout} uses only forecast origins after the public
release of Timestral and serves as a contamination control with low power.

\subsection{Data and carbon intensity}\label{data-and-carbon-intensity}

We use ENTSO-E Transparency Platform data on actual generation per production type (A75) for
the nine Nordic bidding zones NO1--NO5 and SE1--SE4.

For 2019--2025 we use the raw files of Khepri v1.5 unchanged. Data for 2026-01-01 to
2026-11-02 were fetched for this study on 2026-11-02 between 07:12 and 07:28 UTC with
entsoe-py 0.8.0; the last hour complete for all nine zones is 2026-11-02 05:00 UTC.

Raw ENTSO-E data are not redistributed; their sha256 sums are stamped with OpenTimestamps.

Hourly CI (gCO2-eq/kWh) is computed with Khepri v1.5 (DOI 10.5281/zenodo.23101127, git tag
v1.5) without study-specific code: AR5 life-cycle emission factors, the ADR-0002 materiality
threshold, pumped-storage hours set to missing, Waste/Other excluded, hourly mean, and the
v1.5 gap filling \texttt{interpolate(limit=6).ffill().bfill()}.

The same filled series is used as model input and as truth, exactly as in v1.5.

The derived CI series is in the repository (\texttt{data/CI\_KHEPRI-TSFM\_v1\_2026-11-02.parquet}, sha256
ab308ae9f49fb9a4229d3ef089bb8b005fd149101a1e9596c9295c3e5cfa95f0).

\subsection{Forecast origins, horizon and splits}\label{forecast-origins-horizon-and-splits}

Forecast origins are daily at 00:00 UTC. Each forecast covers the 96 hours o+1 h \ldots{} o+96 h, and
an origin is used only if the full 96-hour truth is available.

All arms share the same context end: the origin hour, inclusive.

\emph{Retro.} The Khepri v1.5 2025 test exactly as published: history from 2021 (NO) and 2022 (SE)
(v1.5 data years 2021--2024 and 2022--2024), \texttt{train\_end} 2023-12-31 23:00, test period 2025-01-01 to 2025-12-31, giving
361 origins per zone (2025-01-01 to 2025-12-27).

\emph{Holdout.} Origins from 2026-10-16, the day after the Timestral release on 2026-10-15, with
o+96 h ≤ 2026-11-02 05:00 UTC. This gives 14 origins per zone (2026-10-16 to 2026-10-29).

For SARIMA and GBM the v1.5 training rule is applied with test start 2026-10-16, changing
parameters only (Deviation D-06): \texttt{train\_end} 2025-10-15 23:00, history from 2023 (NO) and 2024
(SE) (data years 2023--2025 and 2024--2025), and GBM training origins 2024-10-12 to 2025-10-11.

Because v1.5 reads whole calendar years, the SARIMA training span becomes 2.79 years (NO) and
1.79 years (SE), against 3.0 and 2.0 years in v1.5; preserving the span exactly would have
required a change to the v1.5 forecasting code, which we did not make.

The window 2026-01-01 to 2026-06-29, already used as a final holdout by Khepri v1.4, is not
used.

\subsection{Forecasting arms}\label{forecasting-arms}

All arms are univariate; no covariates are used.

{\def\LTcaptype{none} 
\begin{longtable}[]{@{}
  >{\raggedright\arraybackslash}p{(\linewidth - 2\tabcolsep) * \real{0.5000}}
  >{\raggedright\arraybackslash}p{(\linewidth - 2\tabcolsep) * \real{0.5000}}@{}}
\toprule\noalign{}
\begin{minipage}[b]{\linewidth}\raggedright
Arm
\end{minipage} & \begin{minipage}[b]{\linewidth}\raggedright
Specification
\end{minipage} \\
\midrule\noalign{}
\endhead
\bottomrule\noalign{}
\endlastfoot
Seasonal naive (week) & f(o+h) = y(o+h−168 h), h = 1\ldots96. A reference arm added for this study (D-04); it enters no test. \\
SARIMA & Khepri v1.5, unchanged: SARIMAX (2,0,1)(1,1,0,24), \texttt{enforce\_stationarity=False}, \texttt{enforce\_invertibility=False}, \texttt{fit(disp=False,\ maxiter=50)} once on the training data; per origin \texttt{apply()} on a 45-day window ending at the origin and a 96-step forecast; diurnal persistence substituted and flagged if a forecast fails. \\
GBM & Khepri v1.5, unchanged: LightGBM with 200 estimators, 31 leaves, learning rate 0.05, seeds 0, 1 and 2 averaged; features day of week, month, CI at the origin, 24 lags and horizon; trained once on the last 365 daily origins whose 96-h targets lie at or before \texttt{train\_end}. \\
Timestral-5M & \texttt{mistralai/Timestral-5M} at revision bdd20cfc15a3fea0e4cb5ccef37f8d3ca7aa335f, package \texttt{timestral} 1.0.0 (github.com/mistralai/timestral at a534bdbd244d64f3d9d743f46c7bf6fc330ed4ff), maximum context 8192, 96-step forecast with library defaults. \\
Timestral-209M & \texttt{mistralai/Timestral-209M} at revision 9874d252ab0aae16f7f7c999f98f98d6ad7d8365, as above. \\
Chronos-2 & \texttt{amazon/chronos-2} at revision 29ec3766d36d6f73f0696f85560a422f50e8498c, \texttt{chronos-forecasting} 2.3.2, \texttt{predict\_quantiles} with prediction length 96 and context length 8192. \\
\end{longtable}
}

The TSFM context is the last 8192 hourly values of the same CI series the baselines use, and
the TSFM point forecast is the 0.5 quantile.

All three TSFMs natively provide the same 21 quantile levels 0.01, 0.05, 0.10, \ldots, 0.90, 0.95,
0.99, which form the common quantile set. SARIMA provides them from its Gaussian predictive
distribution (\texttt{get\_forecast}, predicted mean + \(z_q \cdot\) standard error), with the point forecast
unchanged (D-03). GBM and the seasonal-naive arm give no quantiles.

Khepri v1.5 does not store hourly predictions, so a driver repeats the v1.5 forecasting loop line
by line, calls the same v1.5 functions and keeps the predictions; the v1.5 forecasting module
itself is unchanged (D-01). Before registration this driver was validated against stored v1.5
runs: for SE1 the SARIMA per-(origin, day) MAPE differs by at most 0.00036 pp and GBM by at
most 3.6e-15; for NO1 GBM differs by at most 4.4e-16.

The NO1 SARIMA fit is not bit-reproducible (D-02): in the pinned environment it converges after
1 L-BFGS iteration, against 22 in the 2026-10-01 v1.5 run; in the validation before registration
this gave a NO1 retro mean SARIMA MAPE of 2.4596 against 2.4656 in the stored v1.5 rows. The
reproduction table (Section 3.8) reports the final run. Data, code, library versions and BLAS thread count were
ruled out as causes; the cause was not found. We report this rather than adjust the fit.

The environment is pinned by \texttt{uv.lock} (Python 3.12.13). Inference ran on a CPU (Apple M1, 16 GB)
with \texttt{torch.set\_num\_threads(8)} and the Hugging Face hub offline after download.

\subsection{Loss}\label{loss}

The loss per arm, split, zone and origin is the Khepri v1.5 MAPE over the 96 target hours of
the point forecast; hours with \textbar actual\textbar{} ≤ 1e-6 are excluded.

The test series is the zone-averaged loss per origin, the mean over the nine zones.

For probabilistic forecasts (RQ4) we use the quantile-score approximation of the continuous
ranked probability score (CRPS) over the 21 levels, \(\mathrm{CRPS} \approx \frac{2}{21} \sum_k \mathrm{pinball}_{\tau_k}(y, q_k)\),
averaged over the 96 hours, in gCO2-eq/kWh, and the empirical coverage of the central 80 \%
interval {[}q0.10, q0.90{]} and 90 \% interval {[}q0.05, q0.95{]} (Gneiting \& Raftery 2007).

Origins are never back-filled. If any compared arm misses any zone at an origin, that origin
drops out of the comparison, and the numbers of origins per arm are reported.

\subsection{Research questions, hypotheses and tests}\label{research-questions-hypotheses-and-tests}

The research questions are:
RQ1, whether the TSFMs have lower 96-h MAPE than the Khepri v1.5 SARIMA and GBM baselines on the
post-release holdout; RQ2, the same question on the retro test; RQ3, how the TSFMs do in NO4 and
NO5, the two zones where GBM lost to SARIMA in Khepri v1.5 (+2.92 pp and +0.53 pp), reported
descriptively; RQ4, whether TSFM predictive quantiles are better than SARIMA's on the holdout
(CRPS and coverage); and RQ5, whether the TSFM advantage is larger on the retro test than on the
holdout.

\emph{Hypotheses.} H1: on the retro test a TSFM beats SARIMA (lower mean 96-h MAPE) in at least 5 of
the 9 zones, evaluated separately for each TSFM. H2: on the retro test the TSFM advantage over
SARIMA, in percentage points (pp) of MAPE, is smallest in NO5. Because NO5 has by far the lowest
MAPE level (SARIMA 0.72 \% in v1.5), the pp measure favours H2 by scale alone, so the relative
advantage (pp divided by SARIMA MAPE) is reported as a secondary measure. The null of no
difference is reported as a result in its own right.

\emph{Diebold--Mariano tests.} For each TSFM against each baseline we form
\(d_t = \mathrm{loss}(\mathrm{TSFM})_t - \mathrm{loss}(\mathrm{baseline})_t\) on the zone-averaged per-origin losses and apply the
Diebold--Mariano test (Diebold \& Mariano 1995) with the Harvey--Leybourne--Newbold
small-sample correction (Harvey et al.~1997), h = 4, because 96-hour horizons at daily
origins overlap. The long-run variance uses a rectangular kernel over lags 0\ldots3, with Bartlett
weights over the same lags if it is not positive. The statistic is referred to t(n−1),
two-sided. There are 6 tests per split (3 TSFMs × SARIMA and GBM), each at α = 0.05 unadjusted,
as designed; Holm-adjusted p-values across the 6 tests per split are reported alongside as a
secondary safeguard.

RQ1 uses the holdout tests and RQ2 the retro tests. RQ4 applies the same test to the
zone-averaged per-origin CRPS difference TSFM − SARIMA on the holdout.

\emph{Holdout decision rule.} If \textbar mean d\textbar{} is smaller than the minimum detectable effect (MDE,
Section 3.7) the holdout result is ``not determined'', regardless of p, and never read as
equivalence. If \textbar mean d\textbar{} ≥ MDE and p \textless{} 0.05 the result is ``TSFM better'' (d \textless{} 0) or ``TSFM worse''
(d \textgreater{} 0); otherwise ``not significant''.

\emph{Contamination indicator (RQ5).} With the advantage \(A_t = \mathrm{loss}(\mathrm{baseline})_t - \mathrm{loss}(\mathrm{TSFM})_t\)
(zone-averaged), the indicator is \(\overline{A}_{\mathrm{retro}} - \overline{A}_{\mathrm{holdout}}\), with standard error
\(\sqrt{\mathrm{SE}_{\mathrm{retro}}^2 + \mathrm{SE}_{\mathrm{holdout}}^2}\) from HAC (rectangular, lags 0\ldots3) standard errors of each mean. A
value with z = indicator/SE \textgreater{} 1.96 is reported as ``possible contamination''. The indicator also
reflects any genuine change between 2025 and October 2026 and is not a test of contamination.

Per-zone results are descriptive tables of mean loss per arm, zone and split. Analyses not
listed in the preregistration are labelled exploratory.

\subsection{Statistical power}\label{statistical-power}

The MDE was computed before any TSFM run, from the stored Khepri v1.5 per-(origin, day) metric
rows of the 2025 test (D-05), because no 2025 per-origin predictions were stored. The series is
the zone-averaged SARIMA − GBM loss difference over 361 origins; its long-run variance with a
rectangular kernel and h = 4 is \(\sigma^2_{\mathrm{LR}} = 7.6455\) pp\(^2\).

With \(\mathrm{MDE} = (t_{0.975,\,n-1} + t_{0.80,\,n-1}) / \mathrm{HLN}(n,h) \cdot \sqrt{\sigma^2_{\mathrm{LR}}/n}\), where
\(\mathrm{HLN}(n,h) = \sqrt{(n + 1 - 2h + h(h-1)/n)/n}\), \(\alpha = 0.05\) two-sided and power 0.8, the MDE is
2.989 pp for the n = 14 holdout origins and 0.413 pp for the n = 361 retro origins.

The holdout is therefore a contamination control with low power; the retro test carries the
power.

\subsection{Data traps and sensitivity analysis}\label{data-traps-and-sensitivity-analysis}

Data traps are handled exactly as in Khepri v1.5, without new methods.

In the holdout window, however, the share of hours where the v1.5 CI is missing before gap
filling (pumped-storage hours) is large: 54.5 \% in NO2, 41.6 \% in NO3 and 16.7 \% in NO5, with
longest gaps of 111, 45 and 58 hours; in 2025 the shares are 9.5 \%, 4.7 \% and 4.8 \%. NO1, NO4
and SE1--SE4 have no filled hours in either window (D-07).

Since filled hours enter the truth, the preregistration adds sensitivity analysis S1: the same
Diebold--Mariano tests on the zone average over the six zones without filled hours (NO1, NO4,
SE1, SE2, SE3, SE4), reported descriptively next to the primary tests. The zone set was fixed
from the CI data before any TSFM forecast.

A descriptive reproduction table compares our retro mean MAPE for SARIMA and GBM with the
published v1.5 tables per zone (D-08); it enters no test.

\subsection{Deviations}\label{deviations}

All deviations from the design and the preregistration are logged with dates in the deviation
log; the table below summarises them. D-01 to D-08 were made before registration and are part of it; D-09 and D-10 were
made after the \texttt{prereg-v1} tag and concern only the prospective log, not the arms or the
analysis; D-11 (packaging) and D-12 (wording) change no analysis and no number.

{\def\LTcaptype{none} 
\begin{longtable}[]{@{}
  >{\raggedright\arraybackslash}p{(\linewidth - 4\tabcolsep) * \real{0.3333}}
  >{\raggedright\arraybackslash}p{(\linewidth - 4\tabcolsep) * \real{0.3333}}
  >{\raggedright\arraybackslash}p{(\linewidth - 4\tabcolsep) * \real{0.3333}}@{}}
\toprule\noalign{}
\begin{minipage}[b]{\linewidth}\raggedright
ID
\end{minipage} & \begin{minipage}[b]{\linewidth}\raggedright
Deviation
\end{minipage} & \begin{minipage}[b]{\linewidth}\raggedright
Effect
\end{minipage} \\
\midrule\noalign{}
\endhead
\bottomrule\noalign{}
\endlastfoot
D-01 & Prediction-keeping driver around the unchanged v1.5 forecasting code & Validated against stored v1.5 runs (Section 3.4) \\
D-02 & NO1 SARIMA fit not bit-reproducible; cause not found & Pre-registration validation: NO1 retro SARIMA mean MAPE 2.4596 vs.~2.4656 \\
D-03 & SARIMA quantiles from the Gaussian predictive distribution & Point forecast unchanged \\
D-04 & Seasonal naive (week) added as a new reference arm & Enters no test \\
D-05 & MDE from stored v1.5 per-origin metric rows & No 2025 predictions were stored \\
D-06 & Holdout training rule mapped by parameters only & SARIMA training span 2.79 y (NO), 1.79 y (SE) vs.~3.0 and 2.0 y \\
D-07 & Large share of filled pumped-storage hours in the holdout & Sensitivity S1 on six zones without filled hours \\
D-08 & Reproduction table against published v1.5 results & Descriptive \\
D-09 & Bug fix in the prospective-log issuer (chronological splicing of new A75 data) & Prospective log only; analysis and arms unchanged \\
D-10 & Guard in the prospective-log issuer that skips a day's issue while another heavy job runs; such days are logged as lost, never back-filled & Prospective log only; analysis and arms unchanged \\
D-11 & Khepri v1.5 included byte-identically as \texttt{khepri\_v15/} because its tag has no package definition; only import lines changed & Packaging only; RESULTATER byte-identical in a clean environment \\
D-12 & Preregistration wording on training cut-offs corrected: Chronos-2 became public in October 2025 & Wording only; analyses unchanged \\
\end{longtable}
}

\subsection{Prospective log}\label{prospective-log}

From registration onward, a daily job issues all arms for the 00:00 UTC origin with the holdout
training rule and writes each issue with its sha256 and an OpenTimestamps stamp. Lost days are
logged and never back-filled. The log will be analysed at a later date, not yet fixed, with the same
locked \texttt{analysis.py}, after joining the truth from a later A75 pull through the same Khepri v1.5
CI code. It is not part of the results of this paper.

\subsection{Deferred to a later version}\label{deferred-to-a-later-version}

A 32 768-step context, Timestral-28M and Timestral-123M, and covariates are deferred to a later
version and are not part of this study.

\subsection{Licences}\label{licences}

The code is licensed under Apache-2.0 and the derived data under CC-BY-4.0. The Timestral weights
are under Mistral AI's ``Modified MIT License'', which includes a revenue restriction; Chronos-2
is under Apache-2.0.

\section{Results}
\label{sec:results}

Unless stated otherwise, the numbers in this section come from the locked \texttt{analysis.py} (tag
\texttt{prereg-v1}; run with the import change of deviation D-11) applied to the six prediction files; the output is
\texttt{results/RESULTATER\_KHEPRI-TSFM\_v1\_2026-11-02.md}. Every arm produced forecasts for all
nine zones at all 361 retro origins and all 14 holdout origins; no origin was lost and SARIMA
never fell back to diurnal persistence. Loss is the Khepri v1.5 MAPE over the 96 target hours;
the test series is the zone-averaged loss per origin; a negative difference means that the TSFM
has the lower loss.

\paragraph{Decision rule for the holdout.} The minimum detectable effect for the holdout,
computed before any TSFM run, is $\mathrm{MDE} = 2.989$~pp (Section~\ref{methods}). A holdout
difference whose absolute value is below 2.989~pp is reported as \emph{not determined}, never as
equality, whatever its $p$-value.

\subsection{Reproduction of the baselines}
Our rerun of SARIMA and GBM reproduces the published Khepri v1.5 retro means within
$\pm 0.005$~pp in every zone except NO1 SARIMA ($2.4596$ against the published, rounded $2.47$, a
difference of $-0.0104$~pp; against the unrounded stored v1.5 rows, $2.4656$, the difference is
$-0.006$~pp; the fit converges after one L-BFGS iteration in the current
environment, deviation D-02). These reproduction figures come from
\texttt{results/REPRO-V15\_KHEPRI-TSFM\_v1\_2026-11-02.md}; the NO5 note below comes from
\texttt{results/TILLEGG\_KHEPRI-TSFM\_v1\_2026-11-02.md}.

\subsection{RQ2: retro test (2025, 361 origins)}
\begin{table}[t]
\centering
\caption{Retro test: zone-averaged 96-h MAPE (\%), difference TSFM $-$ baseline (pp) and the
Diebold--Mariano statistic with the Harvey--Leybourne--Newbold correction ($h=4$, $t(n-1)$, $n=361$).
All $p<0.0001$, also after Holm adjustment over the six tests.}
\label{tab:retro}
\begin{tabular}{llrrrr}
\toprule
TSFM & baseline & MAPE TSFM & MAPE baseline & diff. (pp) & HLN stat. \\
\midrule
Timestral-5M & SARIMA & 6.253 & 7.561 & $-1.308$ & $-13.30$ \\
Timestral-5M & GBM & 6.253 & 7.347 & $-1.094$ & $-9.86$ \\
Timestral-209M & SARIMA & 6.213 & 7.561 & $-1.347$ & $-12.77$ \\
Timestral-209M & GBM & 6.213 & 7.347 & $-1.134$ & $-9.97$ \\
Chronos-2 & SARIMA & 6.395 & 7.561 & $-1.166$ & $-12.65$ \\
Chronos-2 & GBM & 6.395 & 7.347 & $-0.952$ & $-7.79$ \\
\bottomrule
\end{tabular}
\end{table}

On the 2025 test all three TSFMs have a lower zone-averaged loss than both SARIMA and GBM
(Table~\ref{tab:retro}). The differences range from $-0.952$~pp (Chronos-2 against GBM) to
$-1.347$~pp (Timestral-209M against SARIMA). The retro test is labelled
\emph{contamination not ruled out}: Timestral discloses neither its training data nor a training
cut-off, and most 2025 origins precede the public release of Chronos-2 in October 2025 (deviation
D-12).

\paragraph{H1.} Each TSFM has a lower retro mean loss than SARIMA in 9 of 9 zones
(Table~\ref{tab:zones}); H1 (at least 5 of 9) is supported for all three arms.

\paragraph{H2.} Measured in percentage points, the smallest advantage over SARIMA is in NO5 for
all three TSFMs, so H2 is supported on its primary measure. As stated in the preregistration,
this measure favours H2 by scale alone: NO5 has the lowest MAPE level (SARIMA 0.722~\%). On the
secondary, relative measure the smallest advantage is in SE1 for all three arms; for
Timestral-209M the relative advantage in NO5 is 17.8589~\%, against 13.8126~\% in SE1.

\begin{table}[t]
\centering
\caption{Mean 96-h MAPE (\%) per zone, retro test (descriptive).}
\label{tab:zones}
\begin{tabular}{lrrrrrr}
\toprule
zone & snaive (week) & SARIMA & GBM & Timestral-5M & Timestral-209M & Chronos-2 \\
\midrule
NO1 & 2.969 & 2.460 & 2.307 & 2.108 & 2.120 & 2.131 \\
NO2 & 4.194 & 3.964 & 3.423 & 3.151 & 3.111 & 3.182 \\
NO3 & 8.139 & 6.874 & 6.583 & 5.868 & 5.773 & 5.910 \\
NO4 & 10.104 & 7.242 & 10.158 & 6.171 & 6.133 & 6.339 \\
NO5 & 1.029 & 0.722 & 1.256 & 0.605 & 0.593 & 0.629 \\
SE1 & 14.322 & 11.638 & 10.758 & 10.096 & 10.031 & 10.266 \\
SE2 & 13.293 & 10.416 & 9.568 & 8.920 & 8.965 & 9.028 \\
SE3 & 5.445 & 4.543 & 4.526 & 3.860 & 3.753 & 3.896 \\
SE4 & 22.605 & 20.187 & 17.545 & 15.495 & 15.441 & 16.174 \\
\bottomrule
\end{tabular}
\end{table}

\subsection{RQ3: NO4 and NO5}
In NO4, where GBM lost to SARIMA in Khepri v1.5, the TSFMs have the lowest retro loss
(Timestral-209M 6.133~\%, Timestral-5M 6.171~\%, Chronos-2 6.339~\%, against SARIMA 7.242~\% and
GBM 10.158~\%). In NO5 the TSFMs are also below SARIMA (0.593--0.629~\% against 0.722~\%). On the
holdout, NO4 shows the same order (TSFM 6.984--7.314~\%, SARIMA 7.480~\%, GBM 9.288~\%). The NO5
holdout carries no information about ranking: the Khepri v1.5 CI series for NO5 is constant at
24.0~gCO$_2$-eq/kWh throughout the holdout window, so every arm except GBM (0.047~\%) has a MAPE of
0.000~\%. This is a property of the data in that window, not a result about the models.

\subsection{RQ1: post-release holdout (14 origins)}
\begin{table}[t]
\centering
\caption{Holdout (origins 2026-10-16 to 2026-10-29): zone-averaged 96-h MAPE (\%), difference
TSFM $-$ baseline (pp), HLN statistic ($t(n-1)$, $n=14$), unadjusted and Holm-adjusted $p$, and the
preregistered verdict ($\mathrm{MDE}=2.989$~pp).}
\label{tab:holdout}
\small\setlength{\tabcolsep}{4pt}
\begin{tabular}{llrrrrrrl}
\toprule
TSFM & baseline & TSFM & base & diff. & stat. & $p$ & $p_{\mathrm{Holm}}$ & verdict \\
\midrule
Timestral-5M & SARIMA & 8.415 & 9.332 & $-0.917$ & $-1.98$ & 0.0693 & 0.1385 & not determined \\
Timestral-5M & GBM & 8.415 & 9.223 & $-0.808$ & $-5.00$ & 0.0002 & 0.0012 & not determined \\
Timestral-209M & SARIMA & 8.005 & 9.332 & $-1.326$ & $-3.79$ & 0.0023 & 0.0068 & not determined \\
Timestral-209M & GBM & 8.005 & 9.223 & $-1.218$ & $-17.97$ & $<$0.0001 & $<$0.0001 & not determined \\
Chronos-2 & SARIMA & 8.494 & 9.332 & $-0.838$ & $-1.89$ & 0.0810 & 0.1385 & not determined \\
Chronos-2 & GBM & 8.494 & 9.223 & $-0.729$ & $-4.95$ & 0.0003 & 0.0012 & not determined \\
\bottomrule
\end{tabular}
\end{table}

On the holdout every TSFM again has the lower point estimate, with differences from $-0.729$ to
$-1.326$~pp (Table~\ref{tab:holdout}). All six differences lie within $\pm 2.989$~pp, so by the
preregistered rule all six comparisons are \emph{not determined}, although four of them have
$p<0.05$. With 14 origins the holdout cannot resolve differences of the size seen on the retro
test; it was designed as a contamination control, not as the test that carries the power.

\paragraph{Sensitivity S1.} Restricted to the six zones without filled pumped-storage hours
(NO1, NO4, SE1--SE4), the retro differences are larger ($-1.171$ to $-1.674$~pp, all
$p<0.0001$) and the holdout differences lie between $-0.981$ and $-1.662$~pp. NO5 is not in the S1
set; in the primary zone average it enters as the preregistration requires, but on the holdout it
contributes a zero loss for every arm except GBM and so carries no information about ranking.

\subsection{RQ4: calibration (holdout)}
\begin{table}[t]
\centering
\caption{Holdout calibration: CRPS (quantile-score approximation over 21 levels,
gCO$_2$-eq/kWh) and empirical coverage of the central 80~\% and 90~\% intervals.}
\label{tab:calib}
\begin{tabular}{lrrr}
\toprule
arm & CRPS & coverage 80~\% & coverage 90~\% \\
\midrule
SARIMA & 1.4165 & 0.835 & 0.911 \\
Timestral-5M & 1.1114 & 0.838 & 0.915 \\
Timestral-209M & 1.0733 & 0.838 & 0.921 \\
Chronos-2 & 1.1198 & 0.806 & 0.905 \\
\bottomrule
\end{tabular}
\end{table}

RQ4 is the preregistered calibration comparison: a DM-HLN test on the zone-averaged per-origin CRPS
difference TSFM $-$ SARIMA was locked in the preregistration, and the MDE applies to MAPE only.
All three TSFMs have a lower holdout CRPS than SARIMA (differences $-0.3052$, $-0.3432$ and
$-0.2967$~gCO$_2$-eq/kWh for Timestral-5M, Timestral-209M and Chronos-2; HLN statistics
$-11.75$, $-11.23$ and $-8.52$; all $p<0.0001$). The SARIMA quantiles are constructed from the
fitted model's Gaussian predictive distribution (deviation D-03), not taken from a native v1.5
probabilistic forecast, so the difference may partly reflect the SARIMA intervals rather than the
TSFMs alone. Empirical coverage lies between 0.806 and 0.838
for the 80~\% interval and between 0.905 and 0.921 for the 90~\% interval across the four arms
(Table~\ref{tab:calib}); these are 14 origins and are not tested.

\subsection{RQ5: contamination indicator}
The indicator is the TSFM advantage on the retro test minus the advantage on the holdout. It
ranges from $-0.084$~pp (Timestral-209M against GBM) to $0.391$~pp (Timestral-5M against
SARIMA), with $z$ between $-0.68$ and $1.75$. No comparison exceeds the preregistered threshold
$z>1.96$, so none is reported as possible contamination. The indicator is descriptive: it also
absorbs any genuine difference between 2025 and October 2026, and with 14 holdout origins its
standard errors are wide (0.124--0.360~pp).

\section{Software and availability}
\label{sec:software}

\subsection{Summary} Khepri-TSFM is a small, preregistered evaluation package: it fetches
ENTSO-E A75 data, derives hourly carbon intensity with Khepri v1.5, runs three zero-shot TSFMs
and the Khepri v1.5 baselines on a fixed set of 96-hour forecast origins, and applies one locked
analysis script. The repository holds the CI series every arm saw, all forecasts with 21
quantiles, the analysis output and the ledgers that tie every number in this paper to a file
and line.

\subsection{Statement of need} Carbon-aware scheduling needs day-ahead to four-day CI forecasts
per bidding zone, and TSFMs are now offered as drop-in forecasters. A comparison against an
already published baseline is only informative if the baseline is not retuned, the forecast
origins and context are shared across arms, and the analysis is fixed before the TSFM forecasts
exist. Khepri-TSFM provides such a comparison for the nine Nordic zones, with a post-release
holdout as a contamination control.

\subsection{State of the field} TSFM-based CI forecasting tools and live TSFM benchmarks exist
(Section~\ref{related-work}). This study is kept as its own repository rather than contributed to
a general toolkit or benchmark because its object is a frozen, versioned comparison: the
baselines, splits, pinned model revisions and analysis must stay bound to the preregistration
and its timestamps, which a contribution into a moving code base would not preserve.

\subsection{Software design} Four trade-offs shape the package. (i) The Khepri v1.5 SARIMA and
GBM code and hyperparameters are used unchanged; a driver repeats the v1.5 evaluation loop only
to keep the hourly forecasts (deviation D-01). (ii) Khepri v1.5 is vendored byte-identically as
\texttt{khepri\_v15}, because its tag has no package definition (D-11). (iii) All TSFMs get the
same 8192-hour context, the maximum of Chronos-2; longer contexts are deferred to v2. (iv)
Everything runs on a CPU (Apple M1, 16~GB): with a batch of nine zones one forecast call takes a
median of 0.143~s (Timestral-5M), 1.800~s (Timestral-209M) and 0.922~s (Chronos-2), and a SARIMA
fit takes 17.8~s (\texttt{results/LATENS\_KHEPRI-TSFM\_v1\_2026-11-02.md}).

\subsection{Research impact statement} The package reproduces the published Khepri v1.5 baselines and
extends them with probabilistic forecasts; the CI series, all forecasts and the analysis can be
re-run from the repository without ENTSO-E access (\texttt{repro\_from\_ci.py}); a daily
prospective log continues the comparison beyond the holdout. We make no claim beyond the nine
zones and the periods tested.

\subsection{AI usage disclosure} Implementation was assisted by Claude Code (Anthropic) as a tool
under the author's direction; roles, tools and versions are listed in
\texttt{AIUSE\_KHEPRI-TSFM\_v1\_2026-11-02.md}. The author is responsible for all choices and
numbers.

\section*{Data and code availability}
Code, the hourly CI series, all forecasts and the analysis output: Zenodo, DOI 10.5281/zenodo.23881315
(repository github.com/avalyset/khepri-tsfm). Khepri v1.5.0: DOI 10.5281/zenodo.23101127.

\section*{AI use}
Implementation was assisted by Claude Code (Anthropic) as a tool under the author's direction
(\texttt{AIUSE\_KHEPRI-TSFM\_v1\_2026-11-02.md} in the repository). The author is responsible for
all choices and numbers.

{\sloppy\small
\section*{References}
\begin{list}{}{\leftmargin=1.5em \itemindent=-1.5em \itemsep=2pt}
\item Ansari, A.\,F., Shchur, O., K\"uken, J., et al. (2025). Chronos-2: From univariate to universal forecasting. arXiv:2510.15821.
\item Botten Nicolaysen, E. (2026). Khepri v1.5.0. Zenodo. doi:10.5281/zenodo.23101127.
\item Clau\ss, J., Stinner, S., Solli, C., Lindberg, K.\,B., Madsen, H., \& Georges, L. (2019). Evaluation method for the hourly average CO2eq.\ intensity of the electricity mix and its application to the demand response of residential heating. \emph{Energies}, 12(7), 1345. doi:10.3390/en12071345.
\item Cramer, E.\,Y., Ray, E.\,L., Lopez, V.\,K., et al. (2022). Evaluation of individual and ensemble probabilistic forecasts of COVID-19 mortality in the United States. \emph{Proceedings of the National Academy of Sciences}, 119(15), e2113561119. doi:10.1073/pnas.2113561119.
\item Diebold, F.\,X., \& Mariano, R.\,S. (1995). Comparing predictive accuracy. \emph{Journal of Business \& Economic Statistics}, 13(3), 253--263. doi:10.1080/07350015.1995.10524599.
\item Garza, A., Rosillo, R., Mendoza-Smith, R., Salinas, D., Williams, A.\,R., Ashok, A., Goswami, M., \& Ju\'arez, J.\,M. (2026). Impermanent: A live benchmark for temporal generalization in time series forecasting. arXiv:2603.08707.
\item Gneiting, T., \& Raftery, A.\,E. (2007). Strictly proper scoring rules, prediction, and estimation. \emph{Journal of the American Statistical Association}, 102(477), 359--378. doi:10.1198/016214506000001437.
\item Harvey, D., Leybourne, S., \& Newbold, P. (1997). Testing the equality of prediction mean squared errors. \emph{International Journal of Forecasting}, 13(2), 281--291. doi:10.1016/S0169-2070(96)00719-4.
\item Hong, T., Pinson, P., Fan, S., Zareipour, H., Troccoli, A., \& Hyndman, R.\,J. (2016). Probabilistic energy forecasting: Global Energy Forecasting Competition 2014 and beyond. \emph{International Journal of Forecasting}, 32(3), 896--913. doi:10.1016/j.ijforecast.2016.02.001.
\item Leerbeck, K., Bacher, P., Junker, R.\,G., Goranovi\'c, G., Corradi, O., Ebrahimy, R., Tveit, A., \& Madsen, H. (2020). Short-term forecasting of CO2 emission intensity in power grids by machine learning. \emph{Applied Energy}, 277, 115527. doi:10.1016/j.apenergy.2020.115527.
\item Maji, D., Sitaraman, R.\,K., \& Shenoy, P. (2022a). DACF: Day-ahead carbon intensity forecasting of power grids using machine learning. In \emph{Proceedings of the Thirteenth ACM International Conference on Future Energy Systems (e-Energy '22)}, 188--192. doi:10.1145/3538637.3538849.
\item Maji, D., Shenoy, P., \& Sitaraman, R.\,K. (2022b). CarbonCast: Multi-day forecasting of grid carbon intensity. In \emph{Proceedings of the 9th ACM International Conference on Systems for Energy-Efficient Buildings, Cities, and Transportation (BuildSys '22)}, 198--207. doi:10.1145/3563357.3564079.
\item Maji, D., Yang, K., Shenoy, P., Sitaraman, R.\,K., \& Srivastava, M. (2025). CarbonX: An open-source tool for computational decarbonization using time series foundation models. arXiv:2510.01521.
\item Mistral AI (2026). Timestral. github.com/mistralai/timestral, commit a534bdbd244d64f3d9d743f46c7bf6fc330ed4ff.
\item Yan, L., Wang, L., Liu, S., \& Ding, Y. (2025). EnsembleCI: Ensemble learning for carbon intensity forecasting. In \emph{Proceedings of the 16th ACM International Conference on Future and Sustainable Energy Systems (e-Energy '25)}, 208--212. doi:10.1145/3679240.3734630.
\end{list}
}

\end{document}